\documentclass[conference]{IEEEtran}

\usepackage[style=ieee]{biblatex}
\renewcommand*{\bibfont}{\footnotesize} %

\usepackage{amsmath,amssymb,amsfonts}
\usepackage{graphicx}
\usepackage{textcomp}
\usepackage[utf8]{inputenc}
\usepackage[english]{babel}
\usepackage{booktabs}
\usepackage[hidelinks]{hyperref}
\usepackage{nowidow}
\usepackage{bm}
\usepackage[dvipsnames]{xcolor}
\usepackage{algorithm}
\usepackage{algpseudocode}

\newcommand*{\sign}{\ensuremath{\text{sign}}}
\newcommand*{\Tr}{\ensuremath{\text{Tr}}}
\newcommand*{\Amat}{\ensuremath{\bm{A}}}
\newcommand*{\amat}{\ensuremath{\bm{a}}}

\newcommand*{\Xmat}{\ensuremath{\bm{X}}}
\newcommand*{\Dmat}{\ensuremath{\bm{D}}}

\newcommand*{\Kmat}{\ensuremath{\bm{K}}}
\newcommand*{\Smat}{\ensuremath{\bm{S}}}
\newcommand*{\Imat}{\ensuremath{\bm{I}}}
\newcommand*{\Qmat}{\ensuremath{\bm{Q}}}
\newcommand*{\Lambdamat}{\ensuremath{\bm{\Lambda}}}

\begin{document}

\title{A Submatrix-Based Method for Approximate Matrix Function Evaluation in the Quantum Chemistry Code CP2K}

\author{\IEEEauthorblockN{Michael Lass\IEEEauthorrefmark{1}\IEEEauthorrefmark{2}, Robert Schade\IEEEauthorrefmark{1}, Thomas D. Kühne\IEEEauthorrefmark{1}\IEEEauthorrefmark{3}, Christian Plessl\IEEEauthorrefmark{1}\IEEEauthorrefmark{2}}
\IEEEauthorblockA{\IEEEauthorrefmark{1}Paderborn Center for Parallel Computing \hspace{2ex} \IEEEauthorrefmark{2}Department of Computer Science}
\IEEEauthorblockA{\IEEEauthorrefmark{3}Department of Chemistry}

\IEEEauthorblockA{Paderborn University, Warburger Str. 100, 33098 Paderborn, Germany}
\IEEEauthorblockA{ \{michael.lass, robert.schade, thomas.kuehne, christian.plessl \}@uni-paderborn.de}
}

\maketitle

\begin{abstract}
Electronic structure calculations based on density-functional theory (DFT) represent a significant part of today's HPC workloads and pose high demands on high-performance computing resources. To perform these quantum-mechanical DFT calculations on complex large-scale systems, so-called linear scaling methods instead of conventional cubic scaling methods are required. In this work, we take up the idea of the submatrix method and apply it to the DFT computations in the software package CP2K. For that purpose, we transform the underlying numeric operations on distributed, large, sparse matrices into computations on local, much smaller and nearly dense matrices. This allows us to exploit the full floating-point performance of modern CPUs and to make use of dedicated accelerator hardware, where performance has been limited by memory bandwidth before. We demonstrate both functionality and performance of our implementation and show how it can be accelerated with GPUs and FPGAs.

\end{abstract}

\section{Introduction}
\label{sec:intro}
The development of new materials and technological processes, for example, for battery technology, photovoltaic or photocatalytic light harvesting or chemical energy conversion, requires the ab initio simulation of large molecules, surfaces or solids with large unit cells.
Density functional theory (DFT,~\cite{PhysRev.136.B864,PhysRev.140.A1133,lieb83_ijqc24_243}) has emerged as an efficient technique for the description of many physical and chemical situations~\cite{RevModPhys.87.897} in this endeavour.

For systems composed of many thousands or even millions of atoms the cubic scaling conventional algorithms for the solution of the DFT Kohn-Sham problem exceed the available computational resources. Thus, methods with a more favourable scaling in the system size are needed to tackle important research questions. The development of linear scaling methods for DFT usually exploits the concept of nearsightedness~\cite{near} in quantum mechanics and are reviewed elsewhere~\cite{RevModPhys.71.1085,doi:10.1002/wcms.1138}. Density-matrix based methods represent one suitable type of linear scaling DFT methods. This family of methods relies on the sparsity of the Kohn-Sham Hamiltonian which in most situations implies a similar sparsity of the one-particle reduced density matrix. Thus, these methods don't solve the DFT problem for the Kohn-Sham wave functions but instead directly purify the Hamiltonian into the density matrix. At zero temperature, this purification can be expressed with the
matrix sign function.

In this paper, we show how the submatrix method~\cite{Lass2018} can be applied to the calculation of the matrix sign function within density-matrix based linear scaling DFT to yield a novel massively-parallel linear scaling DFT method. The main contribution of our work are:
\begin{enumerate}
  \item the proposal and demonstration of a new linear scaling DFT method based on the submatrix method and the matrix sign function, suitable for canonical and grand canonical ensembles at zero or finite temperature,
  \item the adaptation of the submatrix method to domain-specific matrices stored in a distributed format,
  \item an open source implementation of our method within CP2K, and
  \item an initial exploration of hardware acceleration using GPUs and FPGAs.
\end{enumerate}

Section~\ref{sec:cp2k} describes the mathematical framework of density-matrix based linear scaling DFT and the ab initio molecular dynamics program CP2K~\cite{Hutter2013,Kuehne2020} in which we have implemented the new method. The following Section~\ref{sec:submatrix} introduces the submatrix method in the context of linear scaling DFT and the implementation of the method is described in Section~\ref{sec:submatriximplem}. The accuracy and parallel scaling properties of the new method for representative benchmark cases are evaluated in Section~\ref{sec:submatrixeval}. Finally, Section~\ref{sec:submatrixhardware} explores the possibilities to accelerate the submatrix method with GPUs and FPGAs.

\section{Linear Scaling DFT in CP2K}
\label{sec:cp2k}
CP2K~\cite{Hutter2013,Kuehne2020} is an open-source software package for atomistic simulations, providing support for different modeling and simulation methodologies, such as molecular-dynamics (MD) simulations and Monte Carlo (MC) simulations. Forces between atoms can either be computed using classical force fields or using electronic structure methods such as density functional theory (DFT). In this work, we focus on the DFT implementation \emph{Quickstep}~\cite{VandeVondele2005,Kuehne2020} within CP2K. In the following, we briefly describe the parts of CP2K that are relevant for this work.

\subsection{Density-Matrix Based Linear Scaling DFT}
\label{sec:lsdft}
The work horse of large-scale electronic structure calculations for material design, drug development and many other fields is density functional theory (DFT, \cite{PhysRev.136.B864,PhysRev.140.A1133,lieb83_ijqc24_243}).
Within the Born-Oppenheimer approximation, DFT maps the interacting system of electrons to a fictitious non-interacting system with an additional one-particle potential that depends on the electron density. The exponential complexity of the many-particle problem is hidden in the so-called exchange-correlation functional for which approximations like the local density approximations or gradient-based corrections have been derived. The governing equation in DFT is the Kohn-Sham equation
\begin{equation}
\label{eq:schrodinger}
  \left[-\frac{\hbar^2\nabla^2}{2m_e}+\hat V_{ion}+\hat V_{e-ion}+\hat V_{e,\mathrm{eff}}\right] \psi_i(\vec r\,)=\epsilon_i \psi_i(\vec r\,)
\end{equation}
which is a single-particle Schrodinger equation with an effective potential and determines the one-particle wave functions $\psi_i(\vec r\,)$ and the energy levels $\epsilon_i$. $\hbar$ denotes the reduced Planck constant and $m_e$ the mass of the electron.
The first term on the left-hand side in Eq.~\eqref{eq:schrodinger} represents the kinetic energy of the electron and the following terms the potential energies between ions $\hat V_{ion}$, the potential energy of the interaction between electrons and ions $\hat V_{e-ion}$ and the last term $\hat V_{e,\mathrm{eff}}$ the potential due to the effective interaction between electrons. The wave functions $\psi_i(\vec r\,)$ as complex functions in three-dimensional space have to be represented by a finite set of basis functions $\{f_j(\vec r\,)\}$ which have to be chosen before the simulation. Hence, the possible solutions are limited to wave functions that can be represented as
\begin{equation}
  \psi_i(\vec r\,)=\sum_{j} c_{i,j} f_j(\vec r\,)
\end{equation}
with complex coefficients $c_{i,j}$. Inserting this ansatz in Eq.~\eqref{eq:schrodinger} leads to the generalized eigenvalue problem
\begin{align}
\label{eq:generalizedeigenvalue}
  \Kmat \vec c_i &= \Smat \epsilon_i \vec c_i,
\end{align}
where $\Kmat$ is the Kohn-Sham Hamilton matrix and $\Smat$ the so-called overlap matrix that are defined as
\begin{align}
  \label{eq:H}
  K_{i,j}&=\int d^3\vec r\, f_k^*(\vec r\,)\left[-\frac{\hbar^2\nabla^2}{2m_e}+\hat V_{ion}+\hat V_{e-ion}+\hat V_{e,\mathrm{eff}}\right] f_j(\vec r\,) \\
  \label{eq:S}
  S_{i,j}&=\int d^3\vec r f_k^*(\vec r\,) f_j(\vec r\,).
\end{align}
A useful quantity to describe the state of a quantum system with electrons is the one-particle reduced density matrix $D(\vec r,\vec r\,')$ defined in thermodynamic equilibrium at zero temperature as
\begin{equation}
  D(\vec r,\vec r\,')=\sum_i\Theta(\mu-\epsilon_i) \psi_i(\vec r\,) \psi_i^*(\vec r\,'),
\end{equation}
where $\Theta$ is the Heaviside step function. $\mu$ denotes the chemical potential and determines the energy up to which all energy levels are occupied, i.e., $\epsilon_i\leq \mu$ is occupied by electrons. All energy levels larger then the chemical potential ($\epsilon_i>\mu$) are not occupied by electrons. The corresponding definition as a matrix is~\cite{VandeVondele2012}
\begin{equation}
  \label{eq:Dsign}
  \Dmat=\frac{1}{2}\left( \Imat - \sign\left( \Smat^{-1}\Kmat - \mu \Imat \right) \right) \Smat^{-1}
\end{equation}
in terms of the matrix sign function
\begin{equation}
  \sign(\Amat) = \Amat\left( \Amat^2 \right)^{-1/2}.
  \label{eq:sign}
\end{equation}
The sign function maps all eigenvalues $\lambda_i(\Amat), i=1\dots n$ to
\begin{equation}
  \lambda_i(\sign(\Amat)) = \begin{cases}
    +1,& \text{if } \text{Re}(\lambda_i(\Amat)) > 0\\
    -1,& \text{if } \text{Re}(\lambda_i(\Amat)) < 0
    \label{eq:signev}
\end{cases}
\end{equation}
while leaving the eigenvectors of the matrix unchanged.

The diagonal of the density matrix is the electron density $n(\vec r\,)=D(\vec r,\vec r\,)$ and the off-diagonal elements represent the covalency of the system.
The total energy of the system can then be expressed as
\begin{equation}
  \label{eq:bandenergy}
  E=\sum_{i} \epsilon_i+E_{dc}+E_{ion}=\Tr(\Dmat \Kmat)+E_{dc}+E_{ion},
\end{equation}
where the first term denotes the so-called band-structure energy, $E_{dc}$ accounts for double counting terms and $E_{ion}$ for the nuclear Coulomb repulsion.

The conventional solution of the generalized eigenvalue problem scales cubically with the number of basis functions and, thus, cubically with the number of atoms in the system if atom-centered basis functions are used.
With currently available computational resources cubic scaling DFT calculations can only be performed for up to a few thousand atoms.

For large systems with many thousands or even million of atoms the concept of nearsightedness~\cite{near} in quantum mechanics can be exploited. This concept implies that the sparsity patterns of the density matrix $\Dmat$ and the Hamiltonian matrix $\Kmat$ are expected to be similar for most systems. Hence, if the Hamiltonian matrix $\Kmat$ of a system is sparse, then the resulting density matrix $\Dmat$ is also sparse.
By using short-ranged atom-centered basis functions and  truncating very small matrix elements it is possible to make the Hamiltonian even sparser such that the number of non-zero elements scales only linearly with the number of atoms beyond a certain system size. This situation is called the linear scaling regime. When combining a description in the linear scaling regime with a method that also computes the generalized eigenvalue problem in Eq.~\eqref{eq:generalizedeigenvalue} or the matrix sign function in Eq.~\eqref{eq:Dsign} with linear complexity, an overall linearly scaling method can be realized~\cite{VandeVondele2012}.

\begin{figure}
  \includegraphics[width=\columnwidth]{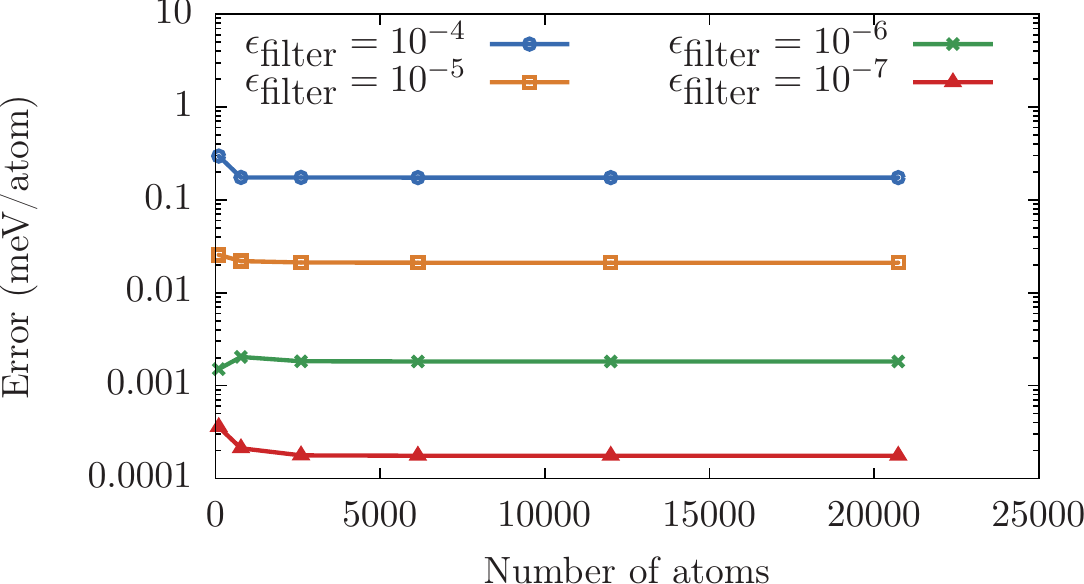}
  \centering
  \caption{Absolute error per atom in total energy computed for systems of liquid water with varying sizes using different truncation thresholds $\epsilon_\text{filter}$. The reference values have been computed with a threshold of $\epsilon_\text{filter}=10^{-12}$. All results have been obtained with a \emph{SZV-MOLOPT-SR-GTH} basis set and using the 2nd-order Newton-Schulz iteration for purification.}
  \label{fig:trunc-error}
\end{figure}

Truncating small matrix values naturally comes at the cost of small errors in the computed results. However, even for increasing system sizes the error relative to the number of atoms stays relatively constant for a fixed truncation threshold (see Figure~\ref{fig:trunc-error} for the error in total energy of exemplary systems of various sizes caused by different truncation thresholds). Related work on ab-initio MD simulations based on approximately calculated forces motivates the use of relatively high truncation thresholds and low-precision arithmetic in the context of electronic structure calculations~\cite{Rengaraj2020}.

\subsection{Iterative Computation of the Sign Function}
\label{sec:sign}
One approach to implement a linear scaling method is to iteratively evaluate the sign function in Eq.~\eqref{eq:Dsign}. Several iterative schemes are available in CP2K, for example the Newton-Schulz iteration\cite{doi:10.1002/zamm.19330130111}
\begin{equation}
  \begin{split}
  \Xmat_0=\Amat,&\quad \Xmat_{k+1}=\frac{1}{2} \Xmat_k(3\Imat-\Xmat_k^2)\\
  \sign(\Amat)&=\lim_{k\rightarrow \infty} \Xmat_k,
  \end{split}
\end{equation}
iterations based on higher-order Padé-approximants~\cite{Higham1997} and arbitrary-order iterations~\cite{Richters2019}.

\subsection{libDBCSR}

\begin{figure}
  \centering
  \includegraphics[width=.7\columnwidth]{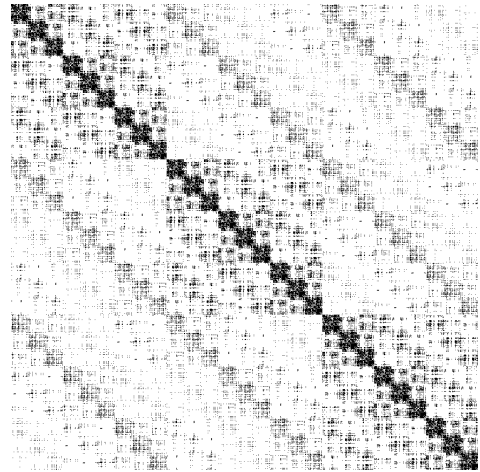}
  \caption{Block-based sparsity pattern of an orthogonalized Kohn-Sham DBCSR matrix $\tilde \Kmat$ for 864 H\textsubscript2O molecules, using an \emph{SZV-MOLOPT-SR-GTH} basis set and a cutoff value of $10^{-5}$, exported from CP2K. Each column corresponds to a water molecule and each black area corresponds to a block that contains at least one non-zero matrix element. }
  \label{fig:szvmatrix}
\end{figure}

A key component of CP2K is the \emph{libDBCSR}~\cite{Borstnik2014} sparse matrix algebra library for distributed computations on large sparse matrices. libDBCSR follows the idea that the sparsity patterns of the processed matrices are not random but typically show certain patterns (see Figure~\ref{fig:szvmatrix} for an example). A matrix stored in DBCSR format is divided into a 2D grid of blocks of relatively small matrices which typically have 5--30 rows and columns. The information which blocks are zero and which contain non-zero elements is stored in CSR format. The blocks with non-zero elements are stored in a dense format. For efficient processing with MPI, DBCSR arranges the MPI ranks in a 2D cartesian topology and creates a mapping from matrix blocks rows and columns to MPI ranks that store these blocks.

libDBCSR provides routines for many matrix operations, in particular matrix-matrix multiplication which is implemented based on a modified Cannon's algorithm~\cite{Cannon1969}. As part of this algorithm, many multiplications of the small DBCSR matrix blocks need to be performed. While this is generally possible using standard BLAS implementations, these are typically not optimized for operation on such small matrices. libDBCSR therefore contains a custom library \emph{libsmm} for small matrix-matrix multiplications. Alternatively, \emph{libxsmm}~\cite{Heinecke2016} can be used for Intel-based systems and there is also a GPU-accelerated version named \emph{libsmm\_acc}~\cite{Schuett2016} (formerly \emph{libcusmm}) included in libDBCSR.

\section{Submatrix Method}
\label{sec:submatrix}
The submatrix method is an algorithm for the computation of approximate solutions to unary matrix functions on large sparse, symmetric matrices. Originally, it has been proposed for the approximate computation of inverse $p$-th roots of matrices~\cite{Lass2018}. In the following, the method is briefly described and its application to the sign matrix function is motivated.

\subsection{Overview}

\begin{figure*}
  \centering
  \includegraphics[width=\linewidth]{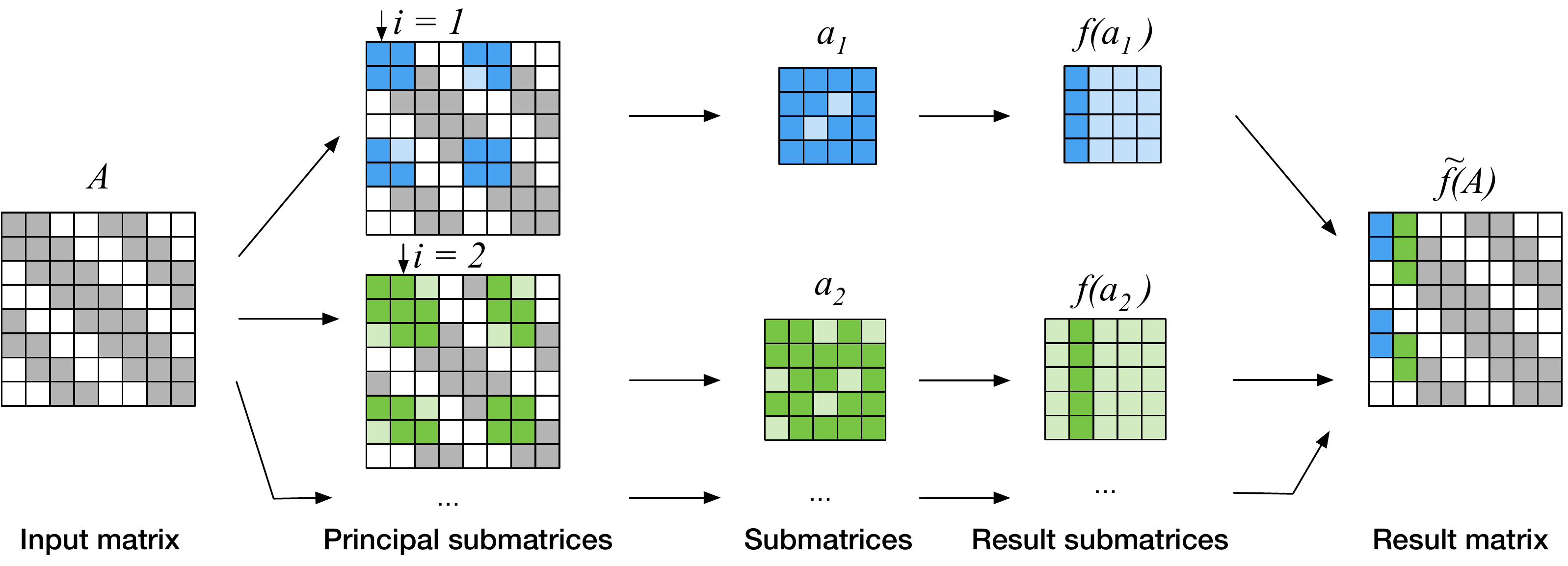}
  \caption{Submatrix Method (adapted from~\cite{Lass2018})}
  \label{fig:sm-scheme}
\end{figure*}

The fundamental idea of the submatrix method is to transform a matrix operation $f$ on a large, sparse $n\times n$ matrix $\Amat$ into $n$ operations on smaller dense matrices. The overall scheme is shown in Figure~\ref{fig:sm-scheme} and can be summarized as follows:

\begin{enumerate}
  \item For each column $i\in 1\dots n$, a principal submatrix $\amat_i$ is assembled by removing all rows and columns $j$ from the original matrix, where $A_{j,i}=0$. The size of the submatrix $\amat_i$ is therefore determined by the number of non-zero elements in the $i$-th column of $\Amat$.
  \item The matrix operation of interest is performed on all submatrices $\amat_i$, resulting in result submatrices $f(\amat_i)$.
  \item Let $k$ be the column within $\amat_i$ that contains the values originating from the $i$-th column of $\Amat$. Then the values from the $k$-th column of $f(\amat_i)$ are used to assemble the $i$-th column of the approximate result matrix $\tilde f(\Amat)$, while retaining the sparsity pattern of the original input matrix $\Amat$.
\end{enumerate}

One major advantage of the method is that it is embarrasingly parallel, since the computation of $f(\amat_i)$ is independent from all other computations $f(\amat_j), j\ne i$. Additionally, the computation of $f$ becomes a dense matrix operation, which often can exploit available computing resources better than sparse matrix operations.

A method similar to the submatrix method has been used in related work~\cite{Niklasson2016,Djidjev2019} where submatrices are determined using graph partitioning algorithms on graphs derived from the sparsity pattern of an existing approximation for the density matrix.

\subsection{Applicability to Sign Function Calculation}
In the original publication of the method~\cite{Lass2018}, the authors evaluate the submatrix method for inverse $p$-th roots of real, symmetric and positive-definite matrices but claim that it is applicable to other operations as well. As shown in Eq.~(\ref{eq:sign}), $\sign(\Amat)$ can be constructed from $\Amat$ and the inverse square-root of $\Amat^2$, suggesting that the sign function is a sufficiently related operation that is also suitable for use with the submatrix method. However, we need to make sure that the sign function is actually defined for all submatrices.

The matrix sign function can be calculated for square matrices which have no eigenvalues on the imaginary axis. In CP2K, the sign function is only applied to square matrices and by construction all submatrices generated as part of the submatrix method are also square.
However, the requirement that no eigenvalues are on the imaginary axis, cannot be guaranteed in CP2K, since the chemical potential $\mu$ can be an arbitrarily chosen in a grand-canonical ensemble and it directly influences all eigenvalues. In CP2K, the definition of the sign function is therefore extended, such that in addition to Eq.~(\ref{eq:signev})
\begin{equation}
  \lambda_i(\sign(\Amat)) = 0,\quad \text{if } \text{Re}(\lambda_i(\Amat)) = 0.
  \label{eq:signevext}
\end{equation}
When using the sign function for computation of the density matrix, this extension is consistent with the physically underlying Fermi function, since
\begin{equation}
  \lim_{\epsilon\rightarrow\mu} \left( \exp\left(\frac{\epsilon - \mu}{k_B T}\right) +1 \right)^{-1} = \frac{1}{2},
\end{equation}
where $k_B$ denotes the Boltzmann constant and $T$ the temperature.
The modification allows us to compute the sign function for all square matrices and therefore also for all submatrices. In our evaluation in Section~\ref{sec:eval}, we will show that the deviation introduced by the submatrix method into chemically relevant quantities like the energy is sufficiently small for our application.

\subsection{Suitability of the Construction for Linear Scaling Methods}

\begin{figure}[!t]
\centering
\includegraphics[width=\columnwidth]{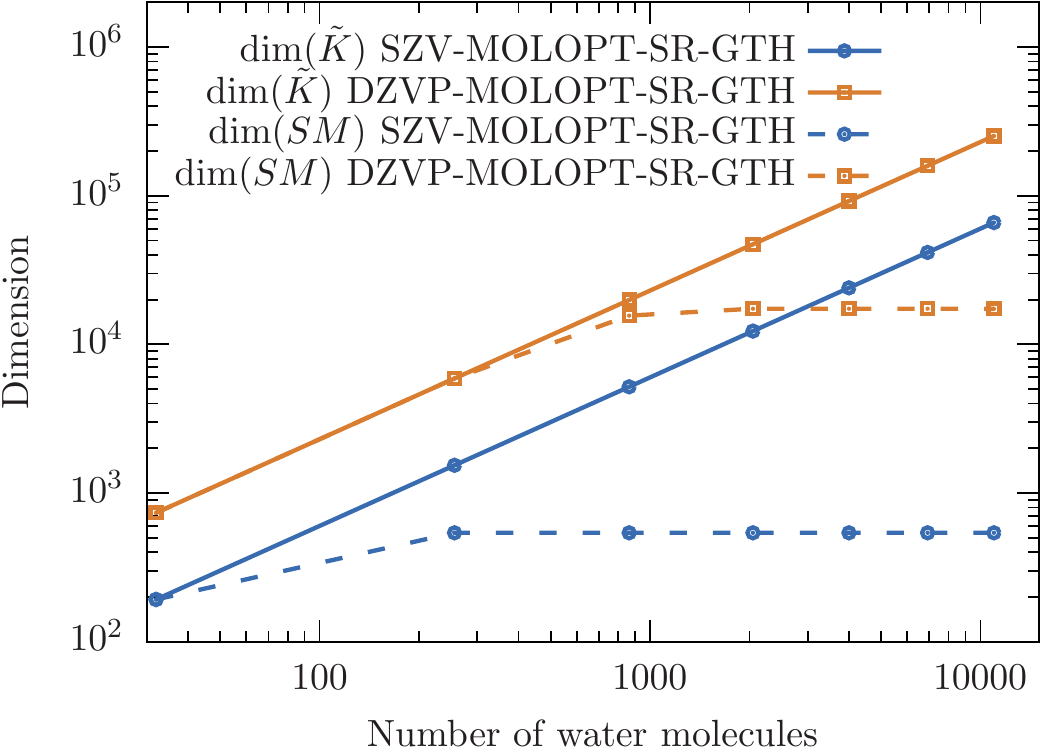}
  \caption{Dimension of submatrices $\mathrm{dim}(SM)$ (block-based, dashed lines) compared to the overall dimension of the orthogonalized Kohn-Sham matrix $\mathrm{dim}(\tilde \Kmat)$ (solid lines) for a cube of liquid water with periodic boundary conditions described in a \emph{SZV-MOLOPT-SR-GTH} (blue, circles) and a \emph{DZVP-MOLOPT-SR-GTH} (orange, cubes) basis set and a cutoff value of $10^{-5}$ for the matrix elements.}
  \label{fig:dim}
\end{figure}

Using atom-centered, short-range basis sets causes matrix elements between basis functions to decay rapidly with the distance of atoms they belong to. Applying a cutoff  $\epsilon_\text{filter}$ to the Hamiltonian to neglect values below a certain threshold results in an overall sparse Kohn-Sham matrix.
Figure~\ref{fig:szvmatrix} shows the symmetrically orthogonalized Kohn-Sham matrix $\tilde \Kmat=\Smat^{-1/2}\Kmat \Smat^{-1/2}$ of a cube of liquid water with 864 H\textsubscript2O molecules, periodic boundary conditions and a single-zeta valence basis set (\emph{SZV-MOLOPT-SR-GTH}). Figure~\ref{fig:dim} compares the dimension of the Kohn-Sham matrix to the dimension of the submatrix for different sizes of a cube of liquid water. Already beyond a total system size of about 200 water molecules the linear scaling regime is reached and the dimension of the submatrices becomes independent of the overall system size which makes the submatrix method a linear scaling method.

The system size beyond which the submatrix method becomes a linearly scaling method depends only on the basis set used and the cutoff of matrix elements. Consider a basis set where the matrix elements for relevant operators such as the Hamiltonian or the overlap operator between basis functions that belong to two different centers decay with the distance of the centers. For a sufficiently large system we can then always determine a finite maximal distance $R_{\mathrm{max}}$ beyond which these matrix elements have absolute values below the cutoff parameter $\epsilon_\mathrm{filter}$. Because the volume defined by the distance $R_{\mathrm{max}}$ is finite and, thus, can only contain a finite number of centers of basis functions, the size of the corresponding submatrix is finite and independent of the overall system size.

\section{Implementation of the Submatrix Method Within CP2K}
\label{sec:submatriximplem}
A basic open source implementation of the submatrix method has been released by the original authors~\cite{Lass2018_github}. This implementation however makes several simplifying assumptions and therefore can only serve as a reference for an implementation within CP2K.

One of these assumptions is that the input matrix is known to all MPI ranks so all of them can create their own submatrices independently. In contrast, in CP2K the matrices are stored in the DBCSR format and therefore in a distributed fashion. Ranks only know about their own blocks of the data. Another difference coming from the DBCSR storage format is that the sparsity of the matrix is only exploited at the level of blocks and not single elements of the matrix. Lastly, the matrices in CP2K have a certain sparsity pattern that depends on the represented chemical system. This pattern needs to be taken into account to minimize data transfers and required floating-point operations and to balance the load between all ranks.

In the following Sections~\ref{sec:impl-overview}--\ref{sec:impl-lb}, we discuss all of these implementation details. Afterwards, we discuss the operation performed on all submatrices in Sections~\ref{sec:diag} and~\ref{sec:mu}.

\subsection{Overview}
\label{sec:impl-overview}
To enable all ranks to assemble their submatrices, a couple of initialization steps need to be performed. Major steps are the following:

\subsubsection{Create Global View on the Sparsity Pattern of the Matrix}
For the input matrix in DBCSR format, each rank only knows which rank is responsible for which blocks of the matrix. However, whether a block is zero or if it contains data is only known to the rank holding that block. To assemble submatrices, each rank needs to know the sparsity pattern of the entire matrix. We achieve this by creating a list of non-zero blocks in a coordinate format (COO), which stores row and column of each non-zero block. This list is deterministically sorted by columns and rows such that it is identical on all ranks. This way, the position of a of non-zero block in this COO representation also serves as a unique ID for the block throughout our implementation.

\subsubsection{Create a Mapping Between Ranks and Submatrices}
The responsibility for creating and processing the submatrices needs to be distributed among all ranks. This happens in a deterministic fashion such that all ranks know which submatrices are solved by which rank. The details of this mapping will be discussed later on.

\subsubsection{Determine Required Matrix Block Transfers}
To assemble a submatrix, the corresponding rank needs a copy of all non-zero blocks that are part of this submatrix. Therefore, we iterate through all blocks of the locally processed submatrices, determine the origin rank and collect the IDs of blocks to be transferred. Additionally, we store a list of all blocks that will be filled with the results calculated locally. These blocks need to be copied back to their origin after finishing the computations.

\subsection{Data Transfers}
\subsubsection{Deduplication of Data Transfers}
In general, data exchange could be implemented as part of the submatrix assembly, such that only blocks required for the currently processed submatrix need to be exchanged and stored.
However, we know that blocks are included in multiple submatrices and these blocks would have to be transferred multiple times between the same ranks. To avoid any duplicate transfers, we make sure that blocks are only transferred once between a pair of nodes by exchanging all required blocks already during the initialization. Each rank stores all blocks that are required for its own submatrices in a local buffer such that submatrices can be assembled without further communication. With this approach we avoid duplicate data transfers and make sure that submatrix assembly becomes a purely local operation.

\subsubsection{Minimization of Memory Use and Data Transfers}
\label{sec:minmem}

To minimize the amount of data that needs to be held in memory and to further reduce the amount of data transfers, ranks should process a set of \emph{similar} submatrices, such that reuse of locally buffered blocks is maximized. Submatrices $\amat_i$ and $\amat_j$ are similar if they share many blocks. This is the case, when columns $i$ and $j$ of the original sparse matrix exhibit a similar sparsity pattern, which depends on the index order of atoms. A similar sparsity pattern of neighboring columns can for example be achieved by sorting the atoms indices to minimize the real-space distances between adjacent indices. For a system constructed of smaller cells of atoms as building blocks the orthogonalized Kohn-Sham matrix usually exhibits a banded structure if the indexing is consecutive in the building blocks. Figure~\ref{fig:szvmatrix} shows an example for the orthogonalized Kohn-Sham matrix for a system of 864 water molecules built up of blocks of 32 water molecules. Thus, in this case a minimization of memory use can be achieved by assigning a consecutive sequence of submatrices to each rank.

\subsection{Minimization of Floating-Point Operations}
To maximize throughput, we also want to minimize the floating-point operations required to obtain a result in addition to data transfer and reuse optimizations. The submatrix method enables a certain trade-off here: While the original idea of the submatrix method is to generate a submatrix for each column of the original matrix, there is also the possibility to generate submatrices from multiple columns.

So far, we apply the submatrix method at the level of DBCSR blocks. This way, we automatically generate submatrices from $b$ consecutive columns, where $b$ is the block width of the corresponding column of the DBCSR matrix. However, there is still the possibility to further split up submatrices to get submatrices for single columns or to combine submatrices built from single block columns.

\subsubsection{Splitting up submatrices}
After assembling a submatrix at the level of DBCSR block columns, it is a regular, densly stored matrix which however still may be sparse. The submatrix method can be applied a second time at the level of single columns to split the submatrix into even smaller, more dense sub-submatrices. Note that it is not required to generate sub-submatrices for all columns of the submatrix. Since submatrix $\amat_i$ only provides values to the overall solution that originate from block column $i$ from the original matrix, it is sufficient to build and solve sub-submatrices for columns originating from block column $i$.

\subsubsection{Generating Submatrices From Multiple Block Columns}
\label{sec:flop-heuristic}
We can also go the other way and combine even more block columns to lower the total number of submatrices $N_S$ that need to be assembled and processed. Because an exhaustive search over all permutations is prohibitively expensive we propose a heuristic based on the assumption that the evaluation of the matrix function for the $i$-th submatrix requires $\mathcal O(n_i^3)$ floating-point operations, where $n_i$ is the dimension of the $i$-th submatrix. That is, the total number of operations is assumed to be
\begin{equation}
  \text{operations} \sim \sum_{i=1}^{N_S} {n_i}^3.
\end{equation}
We define the estimated additional speedup $S$ based on this assumption as
\begin{equation}
  \label{eq:estimatedspeedup}
  S=\frac{\sum_{i=i}^{\tilde N_S} \tilde n_i^3}{\sum_{i=1}^{N_S} {n_i}^3},
\end{equation}
where $\tilde n_i$ are the sizes of the submatrices when generated for single block columns and $\tilde N_S$ the total number of block columns.
The second ingredient are the positions of the atoms $\vec R_i$ in real space that correspond to the $i$-th block-column. When a block column corresponds to more than one atom, we use the center of the set of atoms that corresponds to the block-column. Block columns that should be combined to reduce the number of submatrices are those that correspond to atoms that are close in real space.
Hence, we propose to find a reasonable choice of block columns to combine by applying a clustering algorithm to the positions in real space. %

\begin{figure}
  \includegraphics[width=\columnwidth]{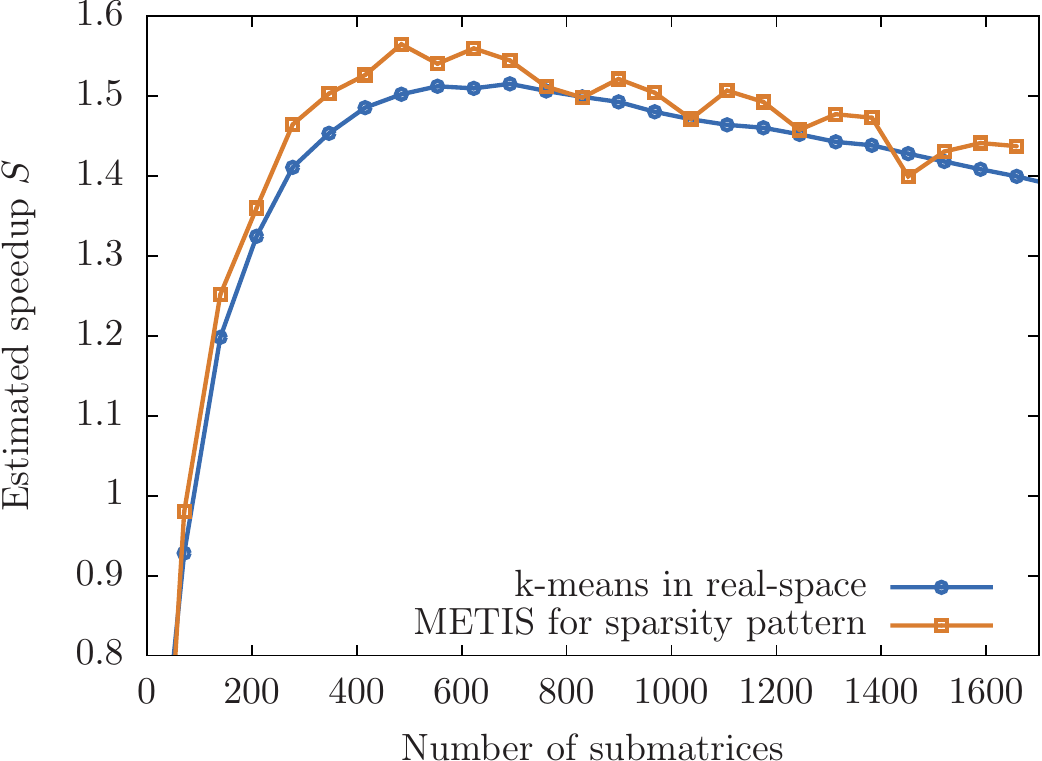}
  \centering
  \caption{Estimated additional speedup S when constructing submatrices from multiple block columns as defined in Eq.~\ref{eq:estimatedspeedup} from a k-means clustering of the real-space coordinates and the clustering of the graph of the sparsity pattern for a system of 6912 water molecules described in an \emph{SZV-MOLOPT-SR-GTH} basis set.}
  \label{fig:clustering-eps}
\end{figure}

The simplest choice of a clustering algorithm is the well-known k-means algorithm~\cite{macqueen1967}. We have used the implementation available in Scikit-learn 0.23.1~\cite{scikit-learn}. Figure~\ref{fig:clustering-eps} shows the dependence of the estimated speedup $S$ on the total numbers of submatrices $N_S$ for k-means-clustering for a system composed of 6912 water molecules with periodic boundary conditions described in an \emph{SZV-MOLOPT-SR-GTH} basis set with a cutoff of matrix elements of $\epsilon_\text{filter}=10^{-7}$.

Another heuristic for combining block columns in submatrices can be derived from the sparsity pattern of the input matrix, i.e., the orthogonalized Kohn-Sham matrix by representing it as a graph. The block columns are the nodes in this graph and there is an edge between two nodes if the corresponding block in the input matrix is not zero. Similar to the approach by Niklasson et al.~\cite{Niklasson2016} we then apply a clustering algorithm to this graph to identify strongly
connected clusters that stand for block-columns that should be combined. The multilevel k-way partitioning from METIS~\cite{METIS} with a total communication volume minimization has turned out to be efficient for this task.

Even though both proposed heuristic are based on very different information, they lead to a similar clustering and with that a similar estimated speedup as shown in Figure~\ref{fig:clustering-eps}. It is surprising that the k-means clustering in real space that only takes $3\cdot 6912$ real numbers as input and doesn't consider the periodicity of the system can compete with the direct clustering of the sparsity pattern that takes a graph with 6912 nodes and 828,540 edges as input in the example shown. This observation shows the viability of the real-space clustering but further investigations are necessary. The proposed heuristic based on the position in real space by construction applies well to spatially inhomogeneous systems of similar molecules or atoms. By adding information about atom types as weights to the clustering problem it can be systematically extended to spatially inhomogeneous systems with different atom species like a virus in a solvent.

For the purpose of this work, the described heuristic has been implemented external to CP2K, leaving an integration into the CP2K software package for near future.

\subsection{Shared-Memory Parallelism}
CP2K supports both distributed memory parallelism and shared memory parallelism. To make use of shared memory parallelism, we use OpenMP to parallelize parts of the initialization of the submatrix method. Routines for the generation of a specific submatrix and for extracting results from a result submatrix are implemented in a thread-safe way, such that these steps as well as solving the submatrices can be implemented using thread-parallelism in the calling code. In our use of the submatrix method for solving the sign function, we distribute the work among all available threads using OpenMP work sharing clauses.

\subsection{Load-Balancing}
\label{sec:impl-lb}
Depending on the chemical system block sizes and sparsity pattern of the DBCSR matrix the submatrix dimensions can vary between different columns of the matrix. For example, a large molecule in solution may have different atom species and exhibit different interactions between its atoms than within the solvent. The matrix columns containing the atoms of the large molecule will therefore induce much larger submatrices. For achieving a good load balance, we therefore cannot just assign the same number of submatrices to each rank but need to consider the estimated computing time to reduce the deviation in execution time between different ranks.

We employ a greedy algorithm to assign submatrices to ranks such that they have similar load. As discussed in Section~\ref{sec:minmem}, we need to find a mapping that assigns one consecutive chunk of submatrices to each rank. Our approach computes the expected number of floating-point operations assuming that processing a submatrix takes $\mathcal O(n^3)$ FLOP and assigns submatrices to ranks as long as their load is expected to be lower than FLOP$_{\mathrm{total}}$/\#ranks. Additionally, we make sure that each rank obtains at least one submatrix.

\subsection{Sign Calculation Based on Diagonalization}
\label{sec:diag}
So far, we have described the implementation of the submatrix method in CP2K. The submatrices can generally be processed using the same mechanism as originally performed on the orthogonalized Kohn-Sham matrix, e.g., by applying a Newton-Schulz iteration scheme. An alternative approach is to use diagonalization to compute the sign function of all submatrices.

To guarantee that the input to the sign function is diagonalizable, we require it to be symmetric. However, although both $\Smat^{-1}$ as well as $\Kmat$ in Eq.~(\ref{eq:Dsign}) are symmetric, their product and therefore the input to the sign function is not. We therefore modify this equation to retain symmetry of the matrices.
The product $\Smat^{-1}\Kmat$ is the orthogonalized Kohn-Sham matrix. Instead of multiplying with $\Smat^{-1}$ we can also apply L{\"o}wdin's symmetric orthogonalization~\cite{lowdin1948quantum} to retain symmetry. We do so by multiplying $\Kmat$ from both sides with $\Smat^{-1/2}$. The density matrix $\Dmat$ can then be computed as
\begin{equation}
  \label{eq:DsignSym}
  \Dmat=\frac{1}{2}\Smat^{-1/2}\left( \Imat - \sign\left( \Smat^{-1/2}\Kmat\Smat^{-1/2} - \mu \Imat \right) \right) \Smat^{-1/2}.
\end{equation}

As shown in Eq.~(\ref{eq:signev}) and~(\ref{eq:signevext}), the sign function can be conceived as an application of the signum function to all eigenvalues. Instead of using iterative schemes, it therefore can be computed using an eigendecomposition of the matrix, for which we use the BLAS routine \texttt{dsyevd} in our implementation:
\begin{align}
  \begin{split}
    \label{eq:eigendecompsym}
    \Amat &= \Qmat\Lambdamat \Qmat^T\\
    \Lambda^\prime_{i,i} &= \text{signum}(\Lambda_{i,i})\\
    \sign(\Amat) &= \Qmat\Lambdamat^\prime \Qmat^T.
  \end{split}
\end{align}

For computing the sign function of our dense submatrices, we found this approach to be superior to iterative approaches. Also, it allows to easily apply our method to systems at finite temperature by replacing the signum function in Eq.~(\ref{eq:eigendecompsym}) by the Fermi function.

\subsection{Adaptation of the Method to Canonical Ensembles}
\label{sec:mu}

As described so far, the submatrix method is a method for grand canonical computations where the chemical potential $\mu$ is fixed, as is the original Newton-Schulz approach. However, solving the submatrices using eigendecompositions allows us to adapt the method also for canonical ensembles, where $\mu$ needs to be dynamically adjusted to compute a density matrix that matches a certain fixed number of electrons.

Using a grand canonical method for a canonical ensemble requires us to compute the total number of electrons as
\begin{equation}
  n_\text{elec} = \Tr\left(\frac{1}{2}\left( \Imat - \sign\left( \Smat^{-1/2}\Kmat\Smat^{-1/2} - \mu \Imat \right) \right)\right)
\end{equation}
and to compare it against the actual number of electrons of the underlying system. If the number of electrons deviate, $\mu$ needs to be adjusted, e.g., using a simple bisection algorithm. Normally this would require recalculation of the sign function in each bisection step, leading to massively increased run times depending on how many bisection steps are required.
Having computed the eigendecomposition of all submatrices allows us to perform this adjustment of $\mu$ without recomputing the sign function or the eigendecomposition in each step, as shown in Algorithm~\ref{alg:mu}.

\begin{algorithm}
\caption{Adjustment of $\mu$ based on eigendecompositions of all submatrices. The single decompositions $\Qmat\Lambdamat\Qmat^T$ only need to be computed once.}
\label{alg:mu}
\begin{algorithmic}
  \State $\mu_\text{corr} \gets 0$
  \Repeat
    \State $n_\text{elec}$ $\gets 0$
    \ForAll{submatrices $\amat^{n\times n}=\Qmat\Lambdamat \Qmat^T$}
      \ForAll{diagonal elements $\lambda_i$ of $\Lambdamat$}
        \State $\lambda_i^\prime = \text{signum}(\lambda_i - \mu_\text{corr})$
      \EndFor
      \ForAll{columns $k$ of $\amat$ that contribute to the \\\quad\quad\quad\quad\quad\quad sparse result matrix}
        \State $n_\text{elec}$ $\gets$ $n_\text{elec} + \frac{1}{2} - \frac{1}{2}\sum_{l=1\dots n} {Q_{k,l}}^2\cdot \lambda^\prime_l$
      \EndFor
    \EndFor
    \State update $\mu_\text{corr}$ based on error of $n_\text{elec}$
  \Until{error of $n_\text{elec}$ is sufficiently small}
  \State $\mu$ $\gets$ $\mu + \mu_\text{corr}$
\end{algorithmic}
\end{algorithm}

Based on the determined value for $\mu$, the sign function for all submatrices can be computed following the scheme from Eq.~(\ref{eq:eigendecompsym}) while adjusting all $\Lambda_{i,i}$ according to the new $\mu$.

In practice, storing all eigendecompositions may be infeasible due to the high memory requirements. However, as shown in Algorithm~\ref{alg:mu}, calculating $\mu$ only requires certain rows from the matrix of eigenvectors $\Qmat$. Most of the additional memory requirements can be saved by only keeping these rows in memory. The downside of this approach is that after determination of the correct value for $\mu$, the submatrices need to be decomposed again in order to compute the final result for the sign function. Still, this approach is superior to recomputing the decomposition in each step of the $\mu$-bisection and we consider it a practical compromise.

\section{Evaluation}
\label{sec:submatrixeval}
\label{sec:eval}
To evaluate our method, we use it within CP2K to compute the density matrix from the Kohn-Sham matrix, following Eq.~(\ref{eq:DsignSym}) with a fixed value for $\mu$, i.e., we consider a grand canonical ensemble. The submatrices are solved using our diagonalization approach, as described in Section~\ref{sec:diag}. For comparison, we look at the default alternative for grand canonical computations which is a 2nd-order Newton-Schulz scheme to compute the sign function of the sparse DBCSR matrix. To make results comparable, we use the same symmetric orthogonalization approach from Eq.~(\ref{eq:DsignSym}), also when using Newton-Schulz iterations.

We perform all computations on a typical benchmark system, which contains liquid water, and use a single-zeta valence basis set (\emph{SZV-MOLOPT-SR-GTH}). The benchmark systems are generated from a fixed-size region containing 32 H\textsubscript2O molecules that is repeated in each dimension by a certain factor \emph{NREP}. The total number of atoms in the system therefore increases with $\text{NREP}^3$.
Due to the fact that the heuristic described in Section~\ref{sec:flop-heuristic} has not been integrated into CP2K yet, submatrices have instead been combined based on a simple greedy heuristic that only considers using a single block column or combining multiples of these basic regions.

For evaluation of the submatrix method, all measurements have been run using a single thread per MPI rank. For measurements of the standard Newton-Schulz method, we used eight ranks per node and five threads per rank, as we found this combination to scale well on our infrastructure.

All measurements in this evaluation have been performed on compute nodes equipped with two Intel Xeon Gold ``Skylake'' 6148 CPUs (40 cores, 2.4 GHz), and 192 GiB of main memory. The nodes are connected by an Intel Omni-Path 100 Gbps network. Program code, input files and raw data of our measurements are publicly available as described in the AD appendix.

\subsection{Performance and Error for Various $\epsilon_\text{filter}$ Thresholds}

\begin{figure}
  \centering
  \includegraphics[width=\columnwidth]{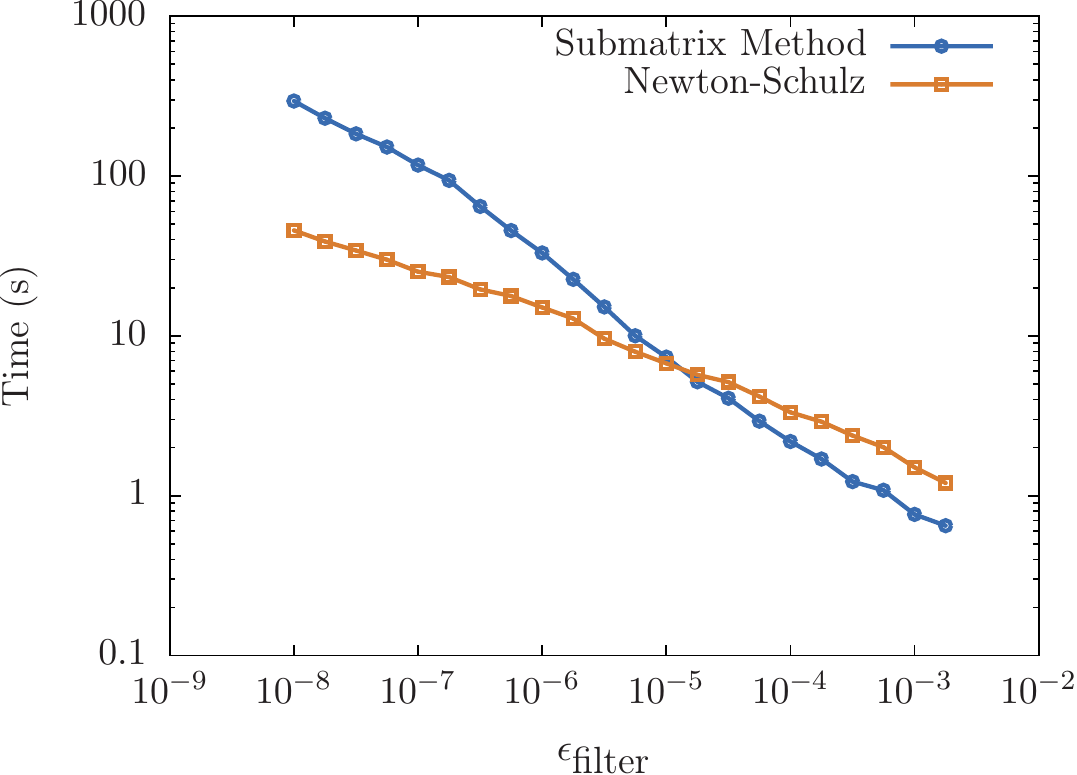}
  \caption{Runtime of submatrix method and 2nd-order Newton-Schulz for various $\epsilon_\text{filter}$ on 80 cores for a system of 20,736 atoms.}
  \label{fig:time-eps}
\end{figure}

To exploit the nearsightedness in quantum mechanics, to obtain sparse matrices and therefore enable linear scaling methods, values below a certain threshold need to be neglected. In CP2K this threshold is called $\epsilon_\text{filter}$ and it is configurable in the input file. For the Newton-Schulz iteration scheme, $\epsilon_\text{filter}$ also determines the convergence criterion. Figure~\ref{fig:time-eps} shows the time required for computation of the density matrix based on different values for $\epsilon_\text{filter}$ for a system where $\text{NREP}=6$ (20,736 atoms) on two compute nodes (80 cores).
The chosen value for $\epsilon_\text{filter}$ significantly influences the run time as for higher values the matrices become more sparse. This effect is even more emphasized for the submatrix method which strongly benefits from the sparsity of the input matrix. For $\epsilon_\text{filter}>10^{-5}$, we observe that the submatrix method becomes quicker than the default Newton-Schulz approach.

\begin{figure}
  \centering
  \includegraphics[width=\columnwidth]{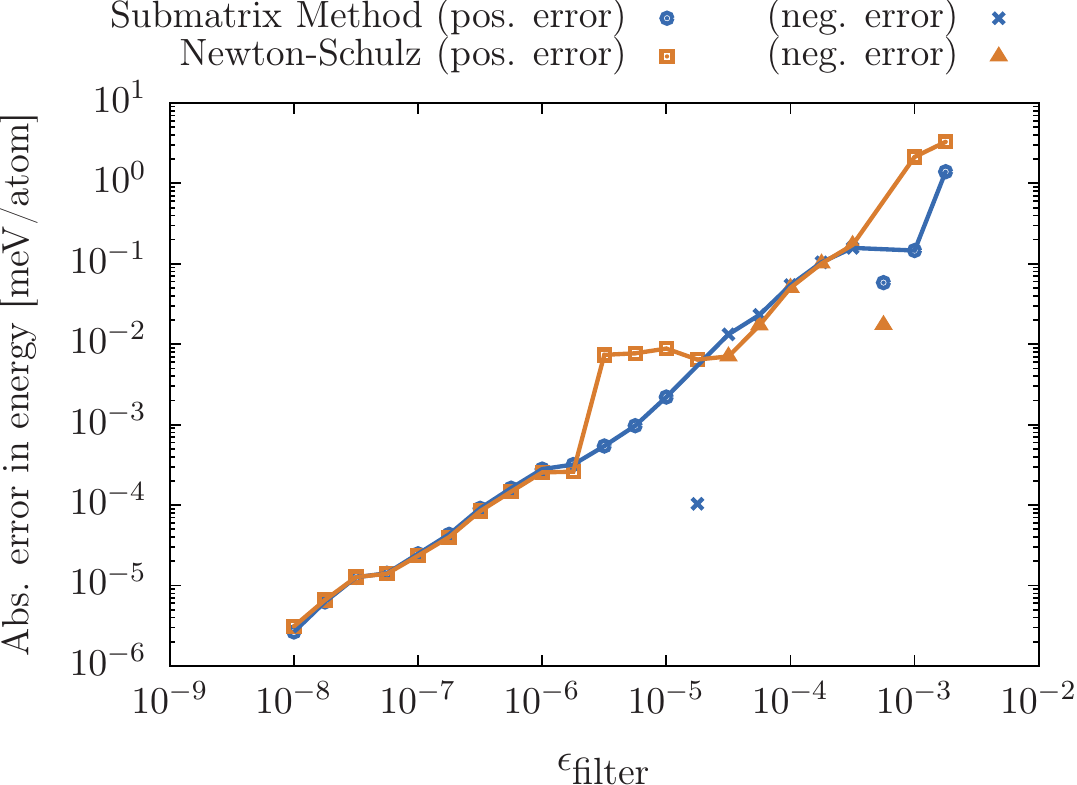}
  \caption{Error in energy computed by the submatrix method and 2nd-order Newton-Schulz for different $\epsilon_\text{filter}$ for a system of 20,736 atoms. The error can be positive or negative, as shown by different markers. The plotted line serves as visual guidance, leaving out datapoints that show a lower absolute error due to the error transitioning between positive and negative values.}
  \label{fig:eps-error}
\end{figure}

Of course, also the resulting error of the cutoff needs to be taken into account. For that we compute the band-structure energy as $\Tr(\Dmat\Kmat)$ (see Eq.~(\ref{eq:bandenergy})) after computation of the density matrix and compare it against a reference value computed with $\epsilon_\text{filter}=10^{-15}$. Results are shown in Figure~\ref{fig:eps-error}. The submatrix method overall shows a resulting error similar to Newton-Schulz which means that the approximation inherent to the submatrix method does not negatively impact the results too much.

\subsection{Scaling}

\begin{figure}
  \centering
  \includegraphics[width=\columnwidth]{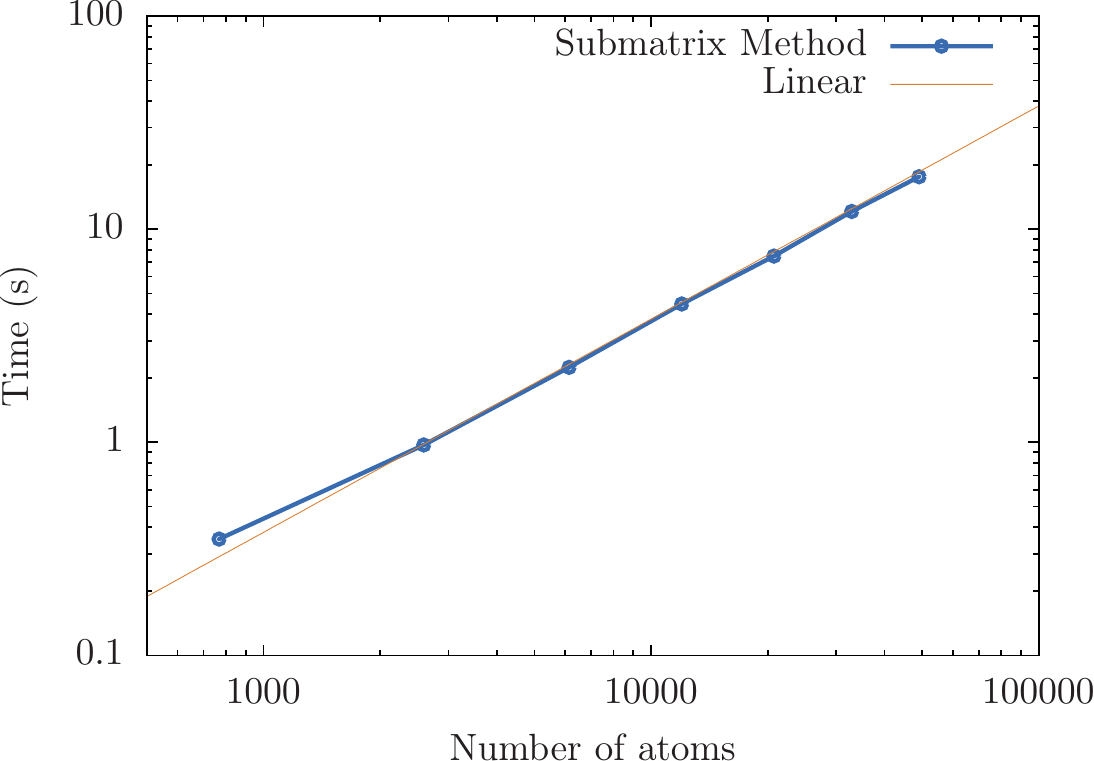}
  \caption{Runtime of submatrix method for increasing system sizes on 80 CPU cores and $\epsilon_\text{filter}=10^{-5}$.}
  \label{fig:linscaling}
\end{figure}

We verify the linear scaling behavior of the submatrix method by scaling up our benchmark system from $\text{NREP}=2$ (768 atoms) to $\text{NREP}=8$ (49,152 atoms) while keeping the amount of computing resources constant at two nodes (80 cores). Results are shown in Figure~\ref{fig:linscaling} and match very well with a linear function.

\begin{figure}
  \centering
  \includegraphics[width=\columnwidth]{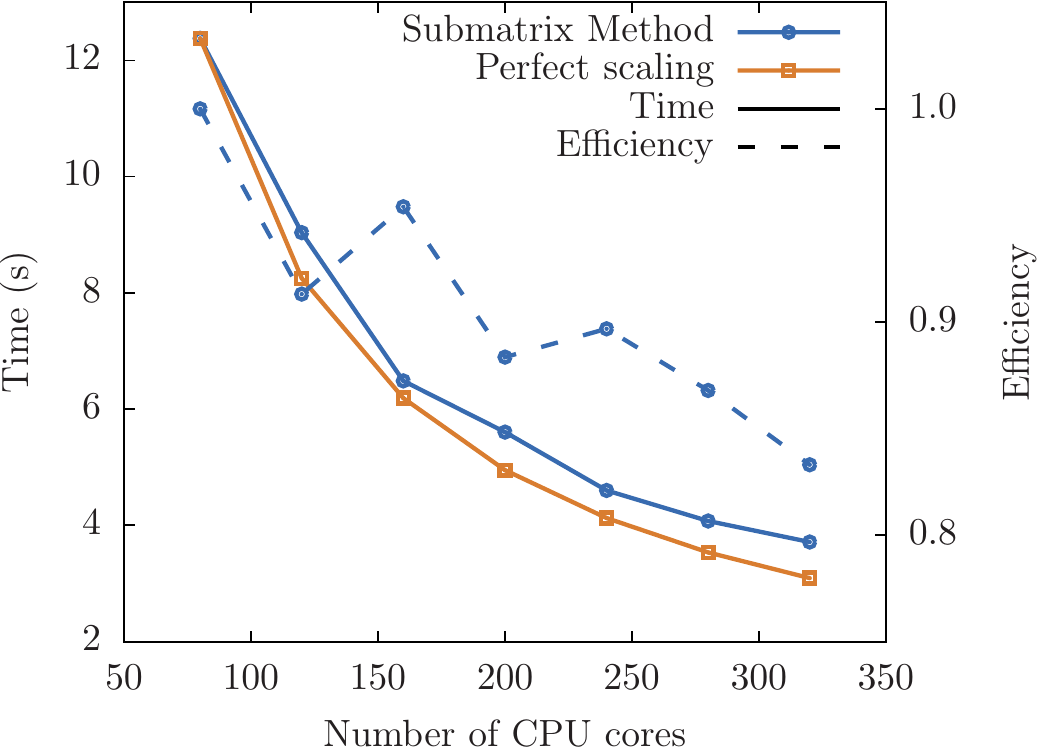}
  \caption{Strong scaling of submatrix method for 32,928 atoms and \mbox{$\epsilon_\text{filter}=10^{-5}$}.}
  \label{fig:strong-scaling}
\end{figure}

To evaluate the strong-scaling behavior of the submatrix method, we take the opposite approach and scale the amount of computing resources between two nodes (80 cores) and eight nodes (320 cores) while keeping the system size fixed at $\text{NREP}=7$ (32,928 atoms). Results are shown in Figure~\ref{fig:strong-scaling}. For comparison, we also show a hypothetical perfect scaling based on the time required on two nodes and the number of nodes used. Going from two to eight nodes, we retain an efficiency of 83\%.

Finally, we evaluate weak scaling, where we increase system size and the amount of compute resources at the same time. To allow more fine-grained control over the system size, we do not replicate the system in all three dimensions but instead use a sufficiently large system of 12,000 atoms ($\text{NREP}=5$) as basis and further replicate it in only one dimension while increasing the number of nodes.

\begin{figure}
  \centering
  \includegraphics[width=\columnwidth]{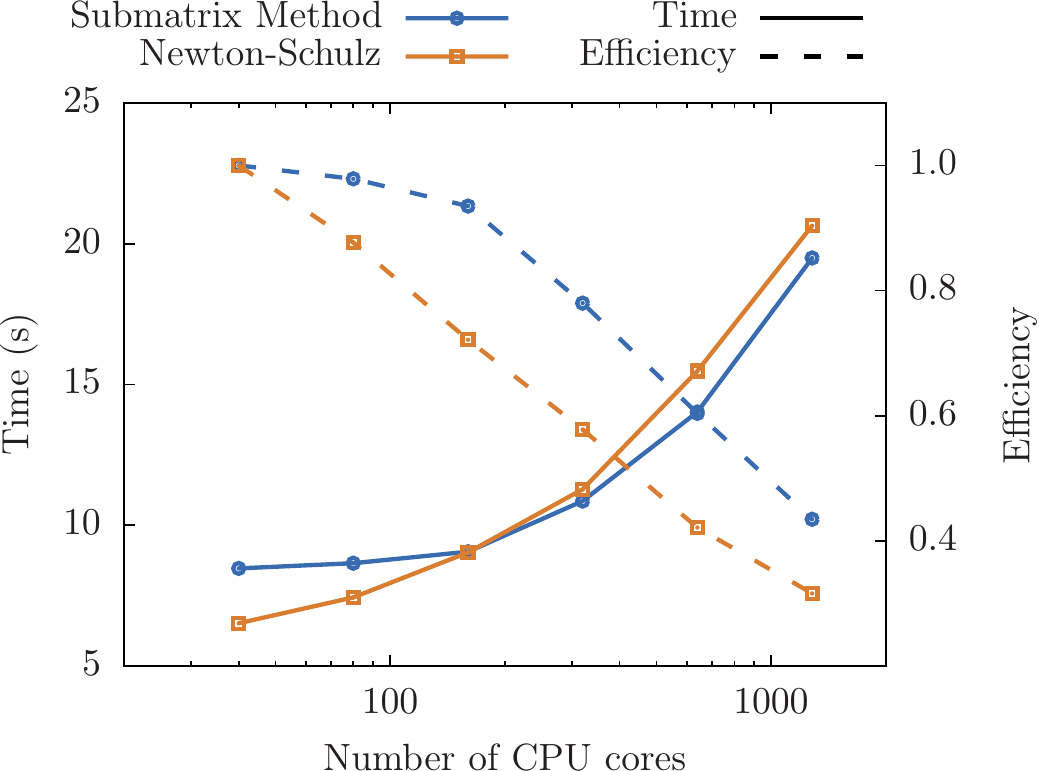}
  \caption{Weak scaling of submatrix method and 2nd-order Newton-Schulz for 12,000--384,000 atoms and 40--1280 CPU cores and $\epsilon_\text{filter}=10^{-5}$.}
  \label{fig:weak-scaling}
\end{figure}

To put the weak-scaling efficiency into perspective, we repeat the same measurements using the standard Newton-Schulz approach, which relies on libDBCSR to scale well over many nodes. Results are shown in Figure~\ref{fig:weak-scaling}. While there is certainly a loss in efficiency when scaling from one to 32 nodes, we see that weak-scaling efficiency is generally higher than for the default Newton-Schulz.

\subsection{Larger Basis Sets}

\begin{figure}[!t]
\centering
\includegraphics[width=\columnwidth]{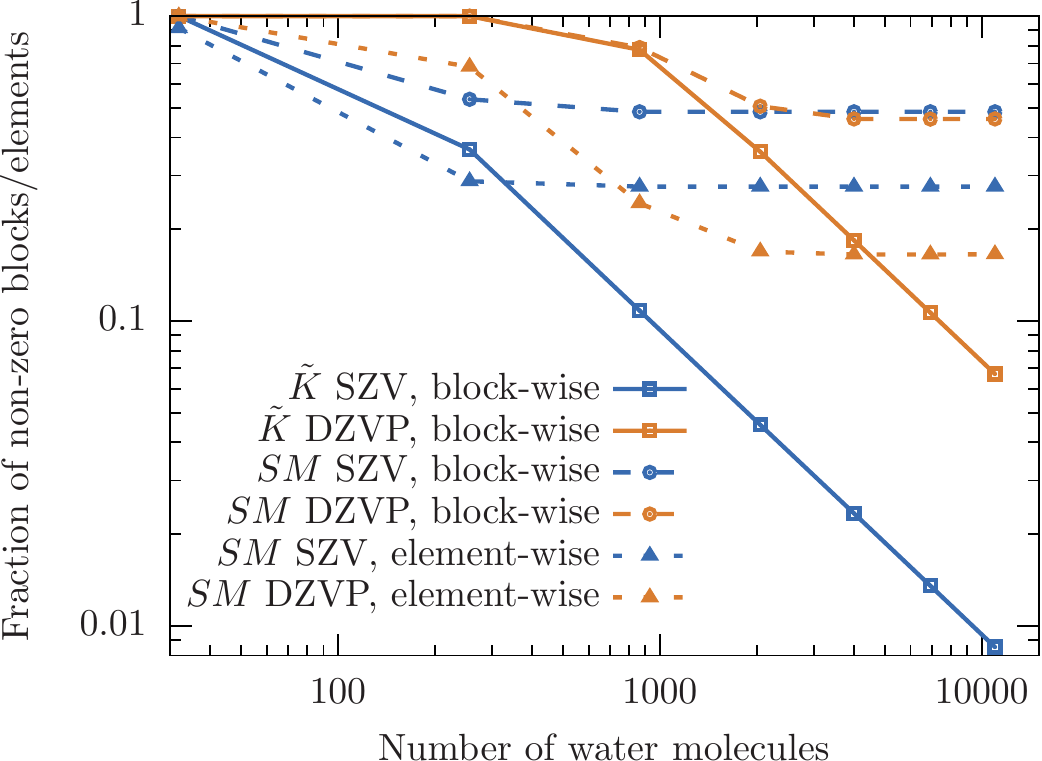}
  \caption{Block-wise (dashed lines with circles) and element-wise (dashed lines with triangles) sparsity of submatrices $SM$ compared to block-wise sparsity the orthogonalized Kohn-Sham matrix $\tilde \Kmat$ (solid lines with circles) for a cube of liquid water with periodic boundary conditions described in a \emph{SZV-MOLOPT-SR-GTH} (denoted as SZV, blue lines) and a \emph{DZVP-MOLOPT-SR-GTH} (denoted as DZVP, orange lines) basis set and a cutoff value of $10^{-5}$ for the matrix elements.}
  \label{fig:sparse}
\end{figure}

Small basis sets like \emph{SZV-MOLOPT-SR-GTH} are often not sufficient for the required chemical accuracy. The submatrix method presented here can also be applied to large basis sets. The main change is that the dimensions of the submatrices grow as shown in Figure~\ref{fig:dim}.
The dimension of the submatrices typically grow stronger than the number of basis states per atom because larger basis sets are usually more long-ranged. In the linear-scaling regime the dimension and sparsity of the submatrices is independent of the system size. When considering the block-wise sparsity as shown in Figure~\ref{fig:sparse} the \emph{DZVP-MOLOPT-SR-GTH} basis set leads to slightly sparser submatrices than the \emph{SZV-MOLOPT-SR-GTH} basis set for a given cutoff value of matrix elements.
In contrast, when an element-wise sparsity is considered, the \emph{DZVP-MOLOPT-SR-GTH} basis leads to much sparser submatrices than the \emph{SZV-MOLOPT-SR-GTH} basis set.
The element-wise sparsities of below 20\% of the submatrices in the \emph{DZVP-MOLOPT-SR-GTH} basis in the linear-scaling regime suggest to replace the evaluation of the sign function of the submatrix that is currently done with dense linear algebra by element-wise sparse linear algebra as a future improvement of the submatrix method especially for larger basis sets.

\section{Hardware Acceleration}
\label{sec:submatrixhardware}
As described in Section~\ref{sec:submatrix} the submatrix method can be seen as an approximate mapping between a matrix function for a large sparse matrix to matrix functions of many smaller but dense matrices. We evaluate here how leveraging accelerators and lower/mixed precision calculations can speed up the evaluation of the sign function for the submatrices and decrease the power consumption for linear scaling DFT.

\subsection{GPU Acceleration Using Tensor Cores}
The tensor cores in modern Nvidia GPUs starting from the Volta-generation can perform matrix-matrix multiplications with a very high energy efficiency. For example, a single consumer-grade Nvidia RTX 2080 Ti can achieve up to 95 TFLOP/s in general matrix-matrix multiplications in half-precision (FP16) at a power efficiency of 380 GFLOP/(Ws). This is possible due to the specialized tensor cores that operate on FP16 input data and perform a 4x4x4 FP16-matrix multiplications in one cycle including either F16 or FP32 accumulation. The Nvidia Cutlass~\cite{cutlass} library as well as Nvidia cuBLAS~\cite{cublas} implement matrix-matrix multiplications of matrices of larger sizes using these tensor core operations. Using the efficient matrix multiplications with tensor cores it is possible the accelerate the sign iterations for submatrices with GPUs. As an example we describe here the GPU implementation of the sign iteration based on the third-order Padé-approximation
\begin{align}
  \begin{split}
  \Xmat_0=\Amat, &\quad \Xmat_{k+1}=\frac{1}{8} \Xmat_k(15-10\Xmat_k^2+3\Xmat_k^4)\\
  \sign(\Amat)&=\lim_{k\rightarrow \infty} \Xmat_k.
  \end{split}
\end{align}
The generalization to different orders is straight-forward. Our implementation uses CUBLAS-routines to program the tensor cores and supports the following precision modes: half precision (FP16), single precision (FP32) and double precision (FP64). Additionally, mixed precision with half-precision multiplication and single-precision accumulation (FP16') is possible. We have used CUDA 10.2 for all evaluations presented here. All steps can be implemented easily with calls to available CUBLAS-functions so that only two matrices have to be transferred from the host to the accelerator.

\begin{figure}[!t]
\centering
\includegraphics[width=\columnwidth]{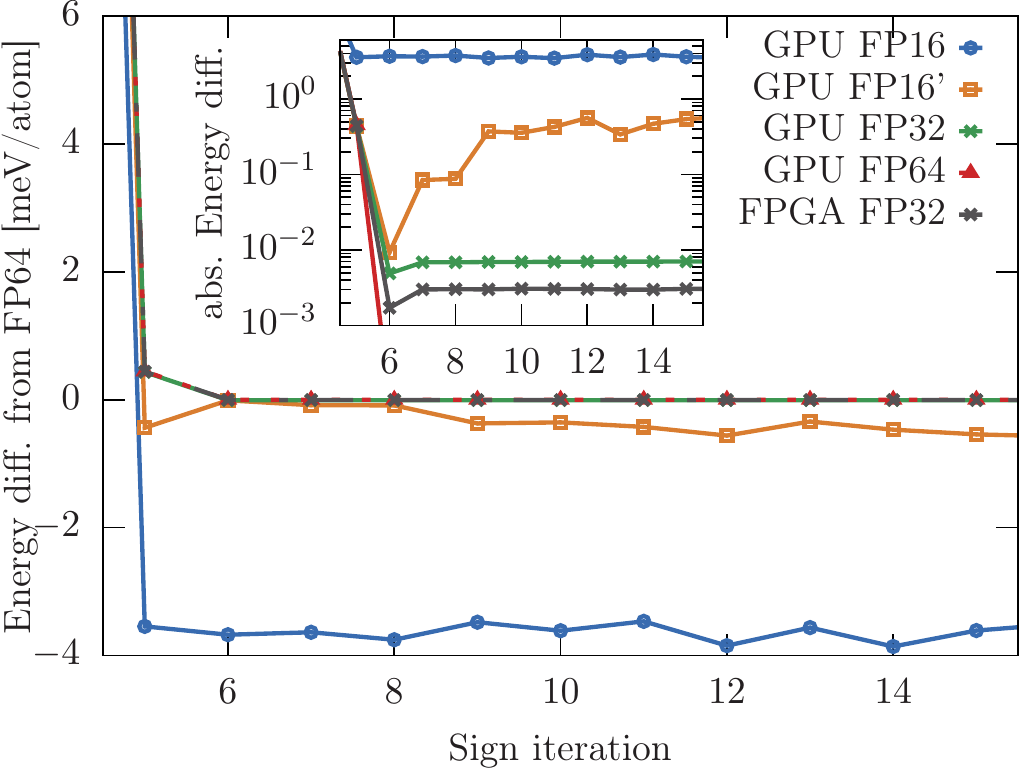}
  \caption{Convergence of the third-order sign iteration in different precisions on a Nvidia RTX 2080 Ti and on a Stratix 10 FPGA for the combined submatrix of 32 water molecules in a system of 4000 water molecules described in an  \emph{SZV-MOLOPT-SR-GTH} basis. The large graph shows the energy difference for the 32 water molecules from the converged FP64 result. The inset shows the absolute energy difference on a logarithmic scale.}
  \label{fig:gpu}
\end{figure}

\begin{figure}[!t]
\centering
\includegraphics[width=\columnwidth]{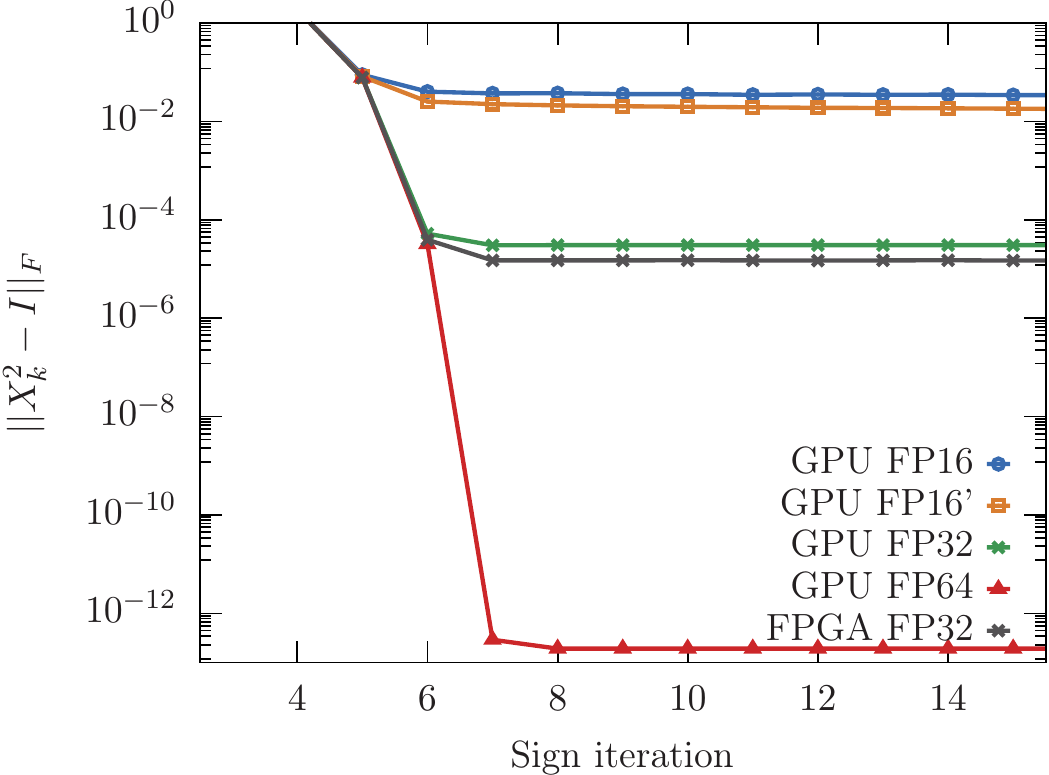}
  \caption{Deviation from the involutority condition $\Xmat_K^2=\Imat$ in every step of the third-order sign iteration in different precisions on a Nvidia RTX 2080 Ti and on a Stratix 10 FPGA for the same situation as in Figure~\ref{fig:gpu}.}
  \label{fig:gpu2}
\end{figure}

Figure~\ref{fig:gpu} shows the convergence of the iteration when executed in different precisions on a Nvidia RTX 2080 Ti for the combined submatrix of 32 water molecules (see Section~\ref{sec:flop-heuristic}) in the same system as in Section~\ref{sec:eval} ($\text{NREP}=5$, total 4000 water molecules, \emph{SZV-MOLOPT-SR-GTH}) .
Interestingly, the resulting energies are within 5 meV/atom of the double-precision result and the sign iteration converges after about 6-8 steps. The inspection of the violation of the involutority condition $\Xmat_{k}^2=I$ in Figure~\ref{fig:gpu2} in every iteration shows that the minimum of the energy is not a suitable convergence criterion because it would signal convergence too early or, in the case of FP16 and FP16', the noise would prevent a detection of convergence.
The theoretical peak performance of a Nvidia RTX 2080 Ti (Turing) in tensor-core-based FP16-matrix-matrix-multiplications for this is about 108 TFLOP/s~\cite{turing} at 250 W energy consumption (432 GFLOP/(Ws)). The size of the submatrix in the present case is 3972 and yields a performance of matrix-matrix multiplies of about 60 TFLOP/s (240 GFLOP/(Ws)). Together with type conversions, transfer to/from the GPU and convergence tests we have achieved an overall practical performance of 35 TFLOP/s at a power consumption of 250 W (140 GFLOP/(Ws)). The results for other precisions are listed in Table~\ref{table:prec}.
\begin{table}[b]
 \centering
  \caption{Peak performance of a Nvidia RTX 2080 Ti~\cite{turing} (second column), practical matrix-matrix multiply performance for the given matrix size 3972 (third column) and overall performance of the sign algorithm including type conversions, data transfer and convergence tests (fourth column) for the different precision modes.}
  \begin{tabular}{ r  r  r  r  }\hline
  precision & peak performance & matrix-multiplies & sign algorithm \\ \hline
    FP16 & 108 TFLOP/s & 56.4 TFLOP/s & 35.2 TFLOP/s \\
    FP16'& 56 TFLOP/s & 38.2 TFLOP/s & 27.8 TFLOP/s \\
    FP32 & 13 TFLOP/s & 12.2 TFLOP/s & 10.4 TFLOP/s \\
    FP64 & 0.5 TFLOP/s & 0.5 TFLOP/s & 0.5 TFLOP/s \\ \hline
  \end{tabular}
  \label{table:prec}
\end{table}

\subsection{FPGA Acceleration of Matrix Multiplications}
FPGAs are a promising architecture for application-specific accelerators. The favourable results of low-precision calculations with GPUs presented in the previous section indicate that single precision or even lower precisions can yield usable results. Thus, we have started to explore the viability of FPGA-designs for the purpose of low-precision calculations for the matrix sign function. As a first step we use existing single-precision matrix-matrix multiply kernels to offload individual matrix multiplications to FPGAs. The studies have been performed with a Bittware 520N board that contains an Intel Stratix 10 GX 2800 FPGA with 32 GB quad-channel DDR4 memory and a PCI-E 3.0 x8 interface.

A few single precision matrix-matrix multiply kernels for Stratix 10 are available~\cite{paolo,fblas}. The Intel FPGA SDK for OpenCL 19.2~\cite{intelfpga} includes an OpenCL-based kernel for matrix-matrix multiplications. We have been able to drastically increase the performance of this kernel with a duplication strategy so that two independent kernels that independently work on $2048\times m \times 2048$ multiplications fit on the board. The resulting design reaches a frequency of 424 MHz and a DSP utilization of 71\%. This results in a practical maximal performance of about 3.4 TFLOP/s for single-precision matrix multiplications at a power consumption of about 110 Watt (31 GFLOP/(Ws)). The blocking of the matrix and the format conversion has to be performed by the CPU.
Due to the fact that at the moment only individual matrix multiplications are offloaded and the PCI-E interface of the Bittware 520N board only uses eight PCI-E lanes, the communication drastically decreases the overall performance. However, the current implementation holds a starting point for more involved application-specific accelerators for the acceleration of the submatrix method.
The convergence of the energy in the sign iterations are shown for the same scenario as for the GPU-based algorithm in Figure~\ref{fig:gpu}. Interestingly, the single-precision results of the GPU and the FPGA don't agree which is most likely due to different blocking of matrix operations which implies a different order of operations. For the submatrix of dimension 3972
in the presented example, %
the Stratix 10 FPGA achieves a practical matrix-multiply performance of 2.7 TFLOP/s in FP32 and an overall sign-function performance of 1.75 TFLOP/s due to the significant communication overhead in the current implementation.

Although the simple FPGA implementation discussed here only achieves about 16 GFLOP/(Ws) compared to about 41 GFLOP/(Ws) for a GPU in single precision, it highlights the importance of avoiding communication by performing all steps of the sign-function algorithm on the device. Furthermore, the flexibility of FPGAs can be used to explore and harness the efficiency of precisions below single or half precision. This however requires devolopment of custom matrix multiplication kernels and needs to take into account efficient use of the FPGA's DSP blocks which for Intel Stratix 10 support variable-precision fixed-point but only single-precision floating-point operations~\cite{intel_dsp}. An evaluation of additional data types on FPGAs therefore remains for future work.

\section{Conclusion}
Density functional theory is a major workhorse in computational chemistry but at the same time highly demanding on compute resources. Scaling up the available resources is costly and generally limited, so development of new methods for efficient DFT computations is essential to allow the application of DFT to larger systems. We have shown that the submatrix method is a promising new method to realize linear scaling DFT computations. Using it in CP2K to compute the density matrix from the Kohn Sham matrix via the matrix sign function, the submatrix method outperforms traditional, iterative approaches if the matrices are sufficiently sparse. At the same time, it exhibits better weak-scaling properties. We have also shown how the submatrix method as an intrinsically grand canonical method can be extended to perform canonical simulations at a small additional computational cost. Finally, we have shown that the method can be generalized to finite temperatures with negligible additional effort.

Due to the fact that the submatrix method transforms the originally sparse and distributed matrix operations to operations on local, dense submatrices allows to approach these operations using new algorithms and new hardware architectures. For the matrix sign function, we have shown hardware acceleration on GPUs and FPGAs, achieving up to 35 TFLOP/s at 250 W using tensor cores. Allowing to exploit these hardware resources, the submatrix method can play an important role in future DFT applications.

By now, our implementation of the submatrix method including its use for computing the density matrix has been included as open source in CP2K. Some of the advanced strategies described in this paper and used in our evaluation are currently being prepared for inclusion. In future, our method and implementation can serve as basis for research and development of more advanced accelerator kernels to further increase efficiency of DFT computations on modern, heterogeneous hardware.

On the other hand, the computation of the full matrix sign function of a submatrix seems wasteful in term of computational efficiency. Thus, efforts are currently on the way that try to selectively calculate selected elements of the sign function for the submatrices in the hopes to improve the overall efficiency of the submatrix method.

\section*{Acknowledgments}
This project has received funding from the German Research Foundation (DFG) under the project PerficienCC (grant agreement No PL 595/2-1).
The authors gratefully acknowledge the funding of this project by computing time provided by the Paderborn Center for Parallel Computing (PC\textsuperscript2).
The support of Tobias Kenter and Paolo Gorlani from PC\textsuperscript2 with running the FPGA tests is gratefully acknowledged.

\printbibliography

\end{document}